**Non-equilibrium Quantum Many-Body Green Function Formalism in the light of Quantum Information Theory**

A. K. Rajagopal [1,2,3]

[1] Inspire Institute Inc., Alexandria, Virginia, 22303, USA

[2] Harish-Chandra Research Institute, Chhatnag Road, Jhunsi, Allahabad 211 019, India

[3] Institute of Mathematical Sciences, C.I.T Campus, Tharamani, Chennai 600 113, India

**Abstract**

The following issues are discussed inspired by the recent paper of Kadanoff (arXiv: 1403:6162 (quant-ph): (1) Construction of a generalized one-particle Wigner distribution (GWD) function (analog of the classical distribution function) from which the quantum kinetic equation due to Kadanoff and Baym(KB) is derived, often called the Quantum Boltzmann Equation (QBE); (2) The equation obeyed by this has a collision contribution in the form of a two-particle Green function. This term is manipulated to have "Kinetic Entropy" in parallel to its counterpart in the classical Boltzmann kinetic equation for the classical distribution function. This proved to be problematic in that unlike in the classical Boltzmann kinetic equation, the contribution from the kinetic entropy term was non-positive; (3) Kadanoff surmised that this situation could perhaps be related to quantum entanglement that may not have been included in his theory. It is shown that GWD is not positive everywhere (indicating "dynamical quantumness") just like the commonly recognized property of the Wigner function (negative property indicating "quantumness" of the state). The issue of non-positive feature appearing in approximate evaluation of patently positive entities in many particle systems is here pointed to an early discussion of this issue (Rajagopal and Sudarshan, Phys. Rev. A10, 1852 (1974)) in terms of a theorem on truncation of cumulant expansion of a probability distribution function due to Marcinkeiwicz. The last issue of presence/absence of entanglement in an approximate evaluation of a many particle correlation poses a new problem; it is considered here in terms of fermionic entanglement theory in the light of density matrix and Green function theory of many-fermion systems. The clue comes from the fact that the Hartree-Fock approximation exhbits no entantanglement in two-particle fermion density matrix and hence also in two-particle Green function.

## 1. Introduction

In a paper entitled "Entropy in Flux", Kadanoff [1] reviewed three ways the quantity "entropy" is introduced in developing various aspects of statistical physics in both equilibrium and non-equilibrium statistical mechanics. Of our interest in this paper is his description of the Kadanoff-Baym (KB) quantum kinetic equation deduced starting from the non-equilibrium Quantum Many-Body Green function formalism. There are three steps in this development: (1) Construction of a generalized one-particle Wigner distribution (GWD) function (analog of the classical distribution function) from which the quantum kinetic equation (KB) is derived, often called the Quantum Boltzmann Equation (QBE); (2) The equation obeyed by this has a collision contribution in the form of a two-particle Green function. This term is manipulated to have "Kinetic Entropy" in parallel to its counterpart in the classical Boltzmann kinetic equation for the classical distribution function. This proved to be problematic in that unlike in the classical Boltzmann kinetic equation, the contribution from the kinetic entropy term was non-positive; (3) He surmised that this situation could perhaps be related to quantum entanglement that may not have been included in his theory.

We first show here that GWD is non-positive in the same fashion as the conventional Wigner distribution associated with a density matrix is non-positive in general (indicating Quantumness inherent in the state) except for Gaussian states. We may attribute "Dynamical Quantumness" to the non-positive property of GWD. Kadanoff assumed the GWD to be real and non-negative in developing the collision terms arising out of approximating a two-particle dynamical Green function. He surmised that this situation could perhaps be related to quantum entanglement that may not have been included in his theory. He cited evidence of such a possibility by referring to [2], where attempts were made to approximate the collision term in several ways. The authors of [2] cited situations of negativity of patently positive entities arising in approximate evaluation of similar types of terms. They attribute this lacuna to possibly not including quantum entanglement in such approximate calculations of the two-particle Green function. The basic source of this has been discussed before [3] and will be presented here in a form related to Schwinger's generating functional of the Green functions of all orders. The entropy and entanglement bounds for Fermi systems residing in the two-particle Fermion state was recently developed in [4]. The purpose of the present paper is to amalgamate all these ideas by re-examining the GWD and its properties as well as issues concerning the source of negativity of patently positive quantities in approximate computations. We use the time-path method of Schwinger-Keldysh for the non-equilibrium one-particle Green function described masterfully in [5] in the development presented herein. This is in the same spirit as the paper of Kadanoff [1]. It should be remarked that there are several types of time path ordering corresponding to different physical situations which define corresponding different types of Green functions discussed in [5]. In the thermal equilibrium case, at absolute zero of temperature, the Green function becomes expectation value of operators defined along corresponding time-ordered path. Thus it is found a direct connection with the density matrix formalism.

The paper is divided into three more sections. In section 2, we follow [5, section 2] in setting up the GWD for one-particle Green function and point out an important property of GWD arising out of the definition of the Wigner transform. We relate the GWD property to the statements made in [1, section

4]. In section 3, we briefly restate the mathematical basis of negativity issues in approximate calculations. In section 4, we state the definition of separability/entanglement in terms of the Green functions, to clarify how to understand this concept in the present context of non-equilibrium statistical mechanics based on the Schwinger-Keldysh Green functions. To the best of our knowledge, the Green function theory of many-particle system has not been formulated to incorporate quantum entanglement issues and we give a preliminary formulation of this. Here we focus on the two-particle Green function which is the central quantity in the Kadanoff discussion. One question that arises from this discussion is whether the presence of quantum entanglement residing in the approximate evaluation of the two-particle Green function may reduce the negative value compared to when it is not included in the scheme. In the final section, a summary of the work is given along with suggestions for further research in this area of research.

## 2. Generalized Wigner Distribution function and inherent Dynamical Quantumness

We follow ref[5] in defining one-particle real time contour ordered Green's function in the Heisenberg picture with arbitrary non-equilibrium density matrix $\rho$:

$$iG(xt;x't') = Tr(\rho T(\psi(xt)\psi^+(x't'))) \tag{1}$$

Here the usual field operators $\psi, \psi^+$ are dependent on space-time $x, t$ (spin is ignored here). There are several contour orderings discussed at length but it suffices here to focus on Schwinger-Keldysh form for simplicity of presentation. The main concern in developing the QBE here is the Green function -G $-iG^<(xt;x't') = Tr(\rho(\psi^+(x't')\psi(xt)))$ appropriate to density of excitations and its GWD whose equation of motion is the QBE. It is worth recalling that in the case of thermal Green function, at absolute zero of temperature, this goes over to the ground state average. The construction of GWD involves introduction of center of mass and relative variables:

$$r = x - x', \quad R = (x + x')/2; \quad t = t - t', \quad T = (t + t')/2 \tag{2}$$

and taking Fourier transform over the relative variables. Thus

$$g^<(k,\Omega,R,T) = \iint dr\,dt\, \exp i(\Omega t - k\,r) G^<\left(R + \frac{1}{2}r, T + \frac{1}{2}t; R - \frac{1}{2}r, T - \frac{1}{2}t\right) \tag{3}$$

We also need to consider product form of two functions belonging to two systems A,B:

$$(G_A^< G_B^<)(xt;x't') = \iint dx_1 dt_1 G_A^<(xt;x_1 t_1) G_B^<(x_1 t_1;x't') \tag{4}$$

And examine the following expression

$$Tr(G_A^< G_B^<) = \iint dx\,dt \iint dx_1 dt_1 G_A^<(xt;x_1 t_1) G_B^<(x_1 t_1;xt) \tag{5}$$

Expressing this in terms of Wigner transforms we obtain

$$Tr(G_A^< G_B^<) = \iint \frac{d^3k\, d\Omega}{(2\pi)^4} \iint dR\, dT\, g_A^<(k,\Omega,R,T)\, g_B^<(k,\Omega,R,T) \tag{6}$$

The left hand side of eq.(6) may be interpreted as a measure of propensity of system A in B, as in the traditional Wigner function construction. If the systems A and B belong to non-intersecting spaces, then LHS is zero, and so conclusion in inescapable that $g^<(k,\Omega,R,T)$ is non-positive. In the thermal Green function case, this is similar to the well-known result concerning the usual Wigner function which is non-positive except when it is associated with Gaussian states. This result about $g^<(k,\Omega,R,T)$ appears to be not known. In the next section, we address the issue of the basic reason for getting negative values for patently positive quantities in approximate calculations of correlations.

## 3. Origin of Negativity issues in developing approximate many-particle correlations

We first observe that in both the references [1] and [2], essentially a two-particle Green's function appears in the dynamical equation for the one particle Green function. In fact, as is well known, the Green function equations do not close and one has a hierarchy of equations. To get the required Wigner transform equation for the one-particle Green function in [1], this set has to be closed and the result should be expressed in terms of one-particle Green functions alone. This procedure is approximate even if accurate to some pre-assigned degree of approximation. In [2], the authors cite random phase approximation which is known to give negative value for the pair correlation function. In [3], several such situations have been cited including the one stated above, and in the theories of turbulence and in quantum theory of partial coherence.

We will give a brief description of the main results developed in [3] about the hierarchies encountered in all of these cases. They are really hierarchies of cumulants of an underlying probability distribution of the many-particle system in the general sense and the approximations basically involve closing the hierarchy by employing cumulants of finite order beyond the second. At this point, recall a mathematical theorem in single variable probability theory (originally stated by Marcinkeiwicz in 1939) for one variable that the probability distribution function will violate its positive definiteness if its cumulant generating function (which is the logarithm of the moment generating function) is a polynomial of degree greater than two. This has been generalized to many variables, fields etc. in order to address the situations that are faced in many particle physics and many other areas of research. The generalization includes the Green function hierarchy in terms of quantum density operator replacing probability functional of many variables and the corresponding moment generating functional generating the Green functions of all orders and the cumulant generating functional. One may refer to [3] for many other references on this topic, especially references 8- 10, and in sec.VI of the paper of Martin and Schwinger in 11, therein. Thus, what the authors of both [1,2] are facing is that the approximations employed by them is equivalent to closing hierarchy of Green function equations by using a finite number of cumulants. This is the origin of the nature of the failure of the underlying probability becoming non-positive which in turn is reflected in some patently positive physical quantity exhibiting negative behavior . In [3], we have cited three more representative new works in areas not covered before, where this basic issue raises its head. The implication of the work in [3] thus concerns the meaning of truncation schemes in many areas of physics. (See specifically reference 13 in [3] in this regard.)

In view of the above result, one cannot attribute the negativity noted in [1,2] to failure to include quantum entanglement in the their approximations, because the negative result may be coming from the innate structure of truncation of cumulant hierarchy.  It is therefore useful and necessary to first check the approximate calculations with and without entanglement and yet employing a truncation of cumulants to find if the negative feature is not as severe at best. It means that we need a test for the presence of entanglement contribution to the part that controls the scattering contribution to the KB equation. In the next section, we develop a definition for entanglement at least for the two-particle Green function that appear in the works of [1,2]. We outline a more general discussion of this topic here and postpone a more comprehensive study of this in a separate paper.

**4. Definition of Separability/Entanglement in terms of Schwinger-Keldysh Green functions**

We begin with a short summary of the new work of Carlen and Lieb (CL)[4] which is relevant to our discussion of the two particle Green function of the many-electron system. This summary serves as a short review and introduction to the entanglement in fermionic system without references to the large body of literature on this topic.

The two-particle density matrix of fermions, especially its entanglement is important in relation to two-particle correlations as in [1, 2, 3]. Fermionic wave function in its simplest form is a determinant. The set of determinants span a space and an orthonormal basis can be constructed out of these. The goal of CL is to show that they have the smallest nonzero entropy and entanglement.

There are two definitions of fermion entanglement:

One, (a) entanglement relative to a Slater determinantal state and the difference is called "correlation".

For references to this, see [4]. The bipartite state, $\rho_{12}$ on the direct product of Hilbert spaces $H_1 \otimes H_2$ of single particles, is finitely separable if it can be decomposed as a convex combination of tensor products of one-particle density matrices $\rho_\alpha$ defined in $H_\alpha$, $\alpha = 1,2$: $\rho_{12} = \sum_{k=1}^{n} \nu_k \rho_1^k \otimes \rho_2^k$, $\nu_k$ *are positive and sum to* $1.$ If not, the bipartite state is entangled. The entanglement of formation, $E_f(\rho_{12})$, in this case is zero if and only if $\rho_{12}$ is separable. Any bipartite fermionic state will be entangled according to this definition.

Two, (b) two-particle density matrix is fermionic-separable if and only if it is convex combination of projections onto 2-partcle Slater determinantal states. Otherwise it is fermionic entangled. One thus looks for a measure of entanglement that is positive on all entangled states and zero on fermionic separable states.

CL prove that convex combinations of Slater determinants uniquely minimize the usual entanglement of formation, $E_f(\rho_{12})$, and therefore the excess of this over the Slater value is a faithful measure of fermionic entanglement. This is their central theorem of interest to us and we state it here:

Theorem (CL): Let $\rho_{12}$ be a bipartite fermionic state. Then $E_f(\rho_{12}) \geq \ln(2)$ and there is equality iff $\rho_{12}$ is a convex combination of pure-state Slater determinants; i.e., the state is fermionic separable. In particular, if $\rho_{12}$ Is the two-particle density matrix of an N-particle fermionic state, then the above inequality is true and equality holds iff the state is fermionic separable.

In view of the discussion in [2] with respect to two-particle electron Green function relevant to the Kadanoff's problem posed in [1], we here consider only the case of fermions. There is a direct relationship of density matrix framework to the Green function formalism, when the former is framed in the second quantized language. The hierarchy of density matrices corresponds to the Green function hierarchy as do the fermionic many-particle versions. The difference between the two formalisms among other things is that the Green function theory has been developed to include both equilibrium and non-equilibrium situations and the associated time-path formalism described in a thorough fashion in [5]. There is a large body of work in many-particle systems based on this formalism (one may see references cited in footnote [3]). Only recently we have a different viewpoint on many-body systems based on quantum information theory (QIT). Fundamentally this is a more detailed study of the intrinsic properties of the many-particle wave function. The latter theory has brought new terms such as separability, entanglement, decoherence etc. which are coming into the discussions in the traditional many-body problems, which also probes the structure of many-particle wave functions. One therefore expects the reflection of the properties discovered in QIT to manifest themselves in many-body theory (MBT). Thus the notion of entanglement in relation to correlation in many particle physics, when approximate methods are employed in the Green function approach to physical systems, has now become an important area of research to find in what ways entanglement appears in such schemes. In this section, we give an outline of how one may connect the two theories.

The many-electron systems have been theoretically investigated by both density matrix and Green function techniques and there are a lot of superficial similarities between the two formalisms but they differ in detail as well as breadth of coverage of physical properties. Underlying both of these the many-particle wave function is central from which the corresponding N-particle density matrix [6] and Green's function [7] are defined. They both are anti-symmetric under exchange of any pair of particles and the underlying wave-function is often described by determinantal wave functions constructed out of a complete set of one-particle wave functions. For our present purposes, we suggest the following parallel definitions: The bipartite state, $\rho_{12}$ on the antisymmetric product of Hilbert spaces $H_1 \Lambda H_2$ of single particles, is finitely separable if it can be decomposed as a convex combination of anti-symmetric tensor products of one-particle density matrices $\rho_\alpha$ defined in sec.3:

$$\rho_{12} = \sum_{k=1}^{n} v_k \rho_1^k \Lambda \rho_2^k, \quad v_k \text{ are positive and sum to } 1.$$ If not, the bipartite state is entangled. In the

Green function case, we propose the corresponding form $G_{12}^{<} = \sum_{k=1}^{n} v_k G_1^{<k} \Lambda G_2^{<k}$, $v_k$ are positive and sum to 1. Otherwise, the bipartite state defined by $G_{12}^{<}$ is entangled. (Note that trace over the single particle density is 1 and that suitably defined over one-particle one-particle Green function defined in the above, is N, the number of fermions.)

The simplest two-particle versions of either theory is in terms of a determinant of two one-particle density matrices and two one-particle Green function. This is gives the well-known Hartree-Fock (HF) approximation for studying the one-particle properties of a system which has only two-body interaction. The correlations between particles beyond this is dealt with by including terms in the wave function involving two-particles explicitly, as in the Hylleraas function. We may cite here the recent works on two-electron systems such as Helium atom where the authors [8] could use such complicated wave function to construct the associated density matrix from which the QIT properties such as entanglement could be examined. Use of a single determinant leading to HF approximation is "not-entangled" and entanglement is found when the wave-function is modified to include a linear combination of determinants or a more sophisticated wave function such as Hylleraas. We take this as a hint that in the Green function method, the simplest approximation of using anti-symmetric combination of two one-particle Green functions [7] (because in the above definition of separability, the sum has only one term) gives no entanglement! Here is one example where the approximation is linked to the entanglement status of the system. In the Green function method, one can make the same observation. In the Green function technique, one can go beyond the HF scheme by including higher order terms by invoking higher order Green functions [7] and still keeping the anti-symmetry intact or equivalently vertex function formulation etc. and derive therefrom the two-particle Green function required in the problems considered in [1,2].

From the simple example of HF approximation and the corresponding Green function approximation, the problem faced in the approximation schemes in both [1,2] is now to find a method to test if quantum entanglement is incorporated. As far as we can gather as of now, such an investigation seeking the presence/absence of entanglement in any approximation of the many particle density matrix or Green function does not exist. One idea that seems to be likely feasible is to construct a Witness, W, a Hermitian operator, and test if the approximate Green function or density matrix is entangled or not.

## 5. Summary and Concluding Remarks

It may be of interest to point out that such unphysical situations were recounted in [3] in wide variety of physical problems some time ago and was shown to be a consequence of a mathematical theorem on truncating cumulant expansion (employed in approximations in the theory) of an underlying probability distribution beyond second order. In this paper, we first show that GWD is non-positive in the same fashion as the conventional Wigner distribution associated with a density matrix is non-positive in general (indicating Quantumness inherent in the state) except for Gaussian states. We may attribute "Dynamical Quantumness" to the non-positive property of GWD. We suggest that this aspect of GWD is avoided if one directly considers the density matrix reformulation of this problem. The notion of entanglement for Fermi systems residing in the two-particle Fermion Green function was recently formulated and developed in [4]. We here define the notion of separability/entanglement in terms of the quantum many-body non-equilibrium Green function in analogy to the definition for density matrices in quantum information theory. In the many electron problem, a pure N-fermion determinantal state gives the Hartree-Fock energy as a first approximation and such a pure state is unentangled, by definition. Improvement of the approximation using linear combination of determinantal wave-functions, for example, introduces possibly entanglement while changing the estimate of the energy of the system. The energy operator is then a "witness" to this entanglement property of the approximation.

**Acknowledgements:** AKR thanks Professor T.V. Ramakrishnan for drawing his attention to the Kadanoff paper [1]. Dr. R. W. Rendell and Professor A. R. Usha Devi read the draft of the paper and made useful remarks.